%% file: PBH_WIMP_final_JCAP-revised-2.tex
\documentclass[a4paper, superscriptaddress, nopacs, preprintnumbers, nofootinbib]{article}
\pdfoutput=1

\include{header}

\begin{document}

\title{Novel Constraints on Mixed Dark-Matter Scenarios of \\Primordial Black Holes and WIMPs}

\author[a,b]{Sofiane M. Boucenna}
\author[a,b]{Florian K{\"u}hnel}
\author[a,b,c]{Tommy Ohlsson}
\author[b,d,e]{Luca Visinelli}

\affiliation[a]{Department of Physics,
	School of Engineering Sciences,\\
	KTH Royal Institute of Technology,
	AlbaNova University Center,\\
	Roslagstullsbacken 21,
	SE--106\.91 Stockholm,
	Sweden}
\affiliation[b]{The Oskar Klein Centre for Cosmoparticle Physics,
	AlbaNova University Center,\\
	Roslagstullsbacken 21,
	SE--106\.91 Stockholm,
	Sweden}
\affiliation[c]{University of Iceland, 
	Science Institute, 
	Dunhaga 3, 
	IS--107 Reykjavik, 
	Iceland}
\affiliation[d]{Department of Physics,
	Stockholm University,
	AlbaNova University Center,\\
	Roslagstullsbacken 21,
	SE--106\.91 Stockholm,
	Sweden}
\affiliation[e]{Nordita,
	AlbaNova University Center,
	Roslagstullsbacken 23,
	SE--106\.91 Stockholm,
	Sweden}

\emailAdd{boucenna@kth.se}
\emailAdd{kuhnel@kth.se}
\emailAdd{tohlsson@kth.se}
\emailAdd{luca.visinelli@fysik.su.se}

\abstract{We derive constraints on mixed dark-matter scenarios consisting of primordial black holes (PBHs) and weakly interacting massive particles (WIMPs). In these scenarios, we expect a density spike of the WIMPs that are gravitationally bound to the PBHs, which results in an enhanced annihilation rate and increased indirect detection prospects. We show that such scenarios provide strong constraints on the allowed fraction of PBHs that constitutes the dark matter, depending on the WIMP mass $m_{\chi}$ and the velocity-averaged annihilation cross-section $\svz$. For the standard scenario with $m_{\chi} = 100\,{\rm GeV}$ and $\svz = 3 \times 10^{-26}\,{\rm cm}^{3} / {\rm s}$, we derive bounds that are stronger than \emph{all} existing bounds for PBHs with masses $10^{-12}\,M_{\odot} \lesssim \MB \lesssim 10^{4}\,M_{\odot}$, where $M_{\odot}$ is the solar mass, and mostly so by several orders of magnitude.}

\date{\formatdate{\day}{\month}{\year}, \currenttime}
\maketitle

\section{Introduction}
\label{sec:Introduction}

\noindent Despite the considerable evidence in favor for the existence of non-baryonic dark matter (DM) in the Universe, the nature of the DM continues to remain unknown and its identification is a major challenge in modern physics. To date, all experimental efforts have yielded null or inconclusive results. From the point of view of particle physics, many theories and paradigms have been proposed to model the DM. A stable weakly interacting massive particle (WIMP) is among the best motivated candidates for describing the presence of DM. WIMPs emerge in various extensions of the Standard Model (SM), which aim to solve other theoretical issues. The natural appearance of WIMPs in such models provides a non-empirical support in their favor.

For masses around the GeV-TeV scale, the annihilation cross-section of WIMPs into lighter particles, computed from a thermal production scenario, is typically of the order of what is obtained from the mediation of the weak interaction. Thus, the relic abundance of WIMPs obtained naturally accounts for the present abundance of the DM. Indeed, after the WIMPs have \emph{chemically} decoupled from the primordial plasma at the freeze-out temperature, their abundance is approximately
\be
	\Omega_{\rm WIMP}\.h^{2}
		\simeq
			0.1\,\frac{3 \times 10^{-26}\.{\rm cm}^{3}\.{\rm s}^{-1}}{\svz^{}_{\rm f.o.}}
			\; ,
			\label{eq:omega}
\ee
where $\svz_{\rm f.o.}$ is the thermal average of the velocity and the annihilation cross-section of the DM into lighter particles at freeze-out and $\Omega_{\rm WIMP} \equiv \rW / \rho_{\rm c}$ is a density parameter with $\rW$ being the energy density of WIMPs and $\rho_{\rm c}$ the critical energy density of the Universe.\footnote{Note that $\rho_{\rm c} = 3\.H_{0}^{2} / 8 \pi G$, where $G$ is the gravitational constant, $H_{0}$ is the present value of the Hubble rate, and $h \equiv H_{0} / (100\,{\rm km} / {\rm s} / {\rm Mpc})$. In the following, the Planck mass $\MP = \sqrt{1/ G\,} \simeq 1.221 \times 10^{19}\,$GeV is also used.} Although the number of WIMPs is fixed at the moment of chemical decoupling, an efficient kinetic equilibrium is still maintained for a certain time through the exchange of momentum between WIMPs and lighter particles. Eventually, the \emph{kinetic} decoupling of WIMPs occurs at a certain temperature $\Tk$ and the WIMP velocity distribution is fixed.

The non-trivial interplay between the known weak-scale physics and DM phenomenology allows for concrete model-building possibilities, relating WIMPs to various aspects of physics lying beyond the SM. This leads to links with, e.g., supersymmetry~\cite{Jungman:1995df}, universal extra-dimensions~\cite{Hooper:2007qk}, and various baryogenesis~\cite{Boucenna:2013wba, Cui:2015eba} and neutrino mass models~\cite{Boucenna:2014zba, Restrepo:2013aga}. Perhaps more importantly, the WIMP paradigm is testable, since it provides many direct and indirect detection prospects through recoil off nuclei and annihilation into detectable SM particles. The ever-growing level of sensitivity reached by WIMP detection experiments offers a great hope that WIMPs will be unambiguously discovered in the near future.

One of the main search strategies for WIMPs is the indirect detection of the products of their annihilation in and beyond our Galaxy. The WIMP annihilation rate is proportional to the square of their number density, implying that denser regions offer much higher detection prospects. Such regions include, e.g., the Galactic center and nearby dwarf spheroidal galaxies. Clearly, any physical mechanism, which increases the density of DM particles, is advantageous from the point of view of indirect detection. A compelling example of such physics is offered by primordial black holes (PBHs), namely black holes that have been produced in the very early Universe.

PBHs have received considerable attention, since they were first postulated~\cite{1967SvA....10..602Z, Carr:1974nx}. The interest in them constituting (parts of) the DM~\cite{1975Natur.253..251C} has recently been revived~\cite{Jedamzik:1996mr, Niemeyer:1997mt, Jedamzik:2000ap, Frampton:2010sw, Capela:2012jz, Griest:2013aaa, Belotsky:2014kca, Young:2015kda, Frampton:2015xza, Bird:2016dcv, Kawasaki:2016pql, Carr:2016drx, Kashlinsky:2016sdv, Clesse:2016vqa, Green:2016xgy, Kuhnel:2017pwq, Akrami:2016vrq, Nakama:2017xvq, Deng:2017uwc, Bernal:2017vvn, Green:2017qoa, Kuhnel:2017bvu, Kannike:2017bxn, Kuhnel:2017ofn, Guo:2017njn, Nieuwenhuizen:2017pto} through the discovery of black-hole binary mergers~\cite{Abbott:2016blz, Abbott:2016nmj}. The possible PBH formation mechanisms are very diverse and there is a plethora of scenarios that lead to their formation. All of these have in common that they require a mechanism to generate large overdensities. Often these overdensities are of inflationary origin~\cite{Hodges:1990bf, Carr:1993aq, Ivanov:1994pa}. When re-entering the cosmological horizon, these overdensities collapse if they are larger than a certain medium- and shape-dependent threshold. Here, the case of radiation domination is the most often considered in the literature. Other scenarios for PBH formation exist, such as those where the source of the inhomogeneities are first-order phase transitions~\cite{Jedamzik:1996mr}, Higgs fluctuations during inflation~\cite{Burda:2016mou, Espinosa:2017sgp}, bubble collisions~\cite{Crawford:1982yz, Hawking:1982ga, Deng:2017uwc}, collapse of cosmic strings~\cite{Hogan:1984zb, Hawking:1987bn}, necklaces~\cite{Matsuda:2005ez} or domain walls~\cite{Berezin:1982ur}. We refer to App.~\ref{sec:Primordial-Black--Hole-Formation} for more details. In some regions of the parameter space, the energy density of PBHs $\rho_{\rm BH}$ can describe the observed DM abundance.

In this paper, we are not assuming any specific formation mechanism of PBHs, since we only require that PBHs are formed prior to the kinetic decoupling of WIMPs from the primordial plasma. We assume a scenario in which the DM is mixed of PBHs and WIMPs, i.e., the energy density of DM is given by $\rDM = \rW + \rho_{\rm BH}$. Thus, the fraction of PBHs $f$ is defined as
\be
	f
		\equiv
			\frac{ \rho_{\rm BH} }{ \rDM }
			\; ,
			\label{eq:Fraction}
\ee
so that the corresponding fraction of WIMPs is $\rW = (1 - f )\.\rDM$. Either PBHs or WIMPs could, in principle, account for the total DM abundance in the Universe. However, even though their existence is well motivated for different reasons, there is no evidence that one of them must account for the total DM abundance. Indeed, a mixed DM scenario, consisting of both PBHs and WIMPs, seems more likely. Such a scenario allows the early-formed PBHs to accrete WIMPs around them and seed the formation of a compact dark clump~\cite{Dokuchaev:2002ts, Ricotti:2009bs, Mack:2006gz, Lacki:2010zf, Saito:2010ts, Eroshenko:2016yve}. This would enhance the annihilation of WIMPs in bound orbits around PBHs even for values of $f$ that are many orders of magnitude below unity.

Assuming a certain simplified density profile of the DM in the central region of the clump, various groups studied the signal produced by DM annihilation in these overdensities~\cite{Lacki:2010zf, Saito:2010ts, Eroshenko:2016yve, Zhang:2010cj, Sandick:2010qu, Sandick:2010yd, Sandick:2011zs}. Here, following Ref.~\cite{Eroshenko:2016yve}, we aim to derive the DM density profile from first principles. Furthermore, our goal is to derive reliable bounds on the parameter space of the mixed PBH-WIMP scenario, using as little assumptions as possible.

The rest of this paper is organized as follows. In Sec.~\ref{sec:Dark--Matter-Density-Spikes-from-Primordial-Black-Holes}, we describe the generation of DM density spikes in the presence of PBHs. Next, in Sec.~\ref{sec:Annihilation-Signal-from-Dark--Matter-Density-Spikes}, we investigate the associated enhancement of the annihilation signal. Then, in Sec.~\ref{sec:Results}, we present, discuss, and summarize our results on constraints on the PBH DM fraction. In Sec.~\ref{sec:Conclusions}, we provide our conclusions. Finally, in the appendices, we give details on the PBH formation mechanism of PBHs (App.~\ref{sec:Primordial-Black--Hole-Formation}), the kinetic decoupling (App.~\ref{sec:Kinetic-Decoupling}), and a derivation of the WIMP density (App.~\ref{sec:Derivation-of-the-WIMP-density}).

\section{Dark-Matter Density Spikes from Primordial Black Holes}
\label{sec:Dark--Matter-Density-Spikes-from-Primordial-Black-Holes}

\noindent Once WIMPs have kinetically decoupled from the primordial plasma, they are gravitationally bound to PBHs and form density spikes, where WIMP annihilation might be boosted even to present days. Comparing the expected annihilation signal with data from the {\sc Fermi} telescope, we can constrain the PBH-WIMP parameter space. In this section, we revise and extend the derivation of the density of WIMPs around a PBH, following Ref.~\cite{Eroshenko:2016yve}. We assume the presence of a sufficient amount of WIMPs.
 
In a radiation-dominated Universe, the Hubble rate depends on the cosmic time $t$ and the temperature of the plasma $T$ as
\be
	H(T)
		=
			\frac{1}{2\.t}
		=
			\sqrt{\frac{8\pi\.G}{3}\.\rho\,}
		=
			\frac{\alpha\,T^{2}}{2\.\MP}
			\; ,
			\label{eq:H(T)}
\ee
where $\rho$ is the energy density of radiation and we introduce the quantity
\be
	\alpha 
		\equiv
			\sqrt{\frac{16\pi^{3}\.g_{*}(T)}{45}\,}
			\label{eq:Alpha}
\ee
in terms of the effective number of relativistic degrees of freedom $g_{*}(T)$. For practical purposes, we set $g_{*}(\Tk) = 61.75$. At temperature $\Tk$, the scattering of WIMPs off radiation becomes inefficient in exchanging momentum and WIMPs kinetically decouple from the plasma. From the moment of decoupling, the WIMP momentum $p$ decreases with the scale factor $a(t)$ according to $p \propto 1/a(t)$, which leads to a WIMP temperature:
\be
	T_{\rm WIMP}
		=
			\Tk\,\frac{\tk}{t}
			\; ,
	\qq
	\hbox{for $t \geq \tk$}
			\; .
			\label{eq:TWIMP}
\ee
Note that the behavior of $T_{\rm WIMP}$ differs from the evolution of $T$, which scales as $a^{-2}(t)$. We refer to App.~\ref{sec:Kinetic-Decoupling} for additional details on the derivation of $\Tk$.

Suppose that, at time $t_{i}$ prior the kinetic decoupling $\tk$, a PBH of mass $\MB$ forms in a radiation-dominated Universe with the energy density $\rho$. Using Eq.~\eqref{eq:H(T)}, we have $\rho = 3 / ( 32 \pi\.G\.t^{2} )$. In order for particles to be gravitationally affected by the PBH, the energy density within a sphere of radius $\ri( \tk )$ must equal $\rho$:
\be
	\rho
		=
			\frac{\MB}{\frac{4\pi}{3}\.\ri( \tk )^{3}}
			\; .
			\label{eq:rho}
\ee
This yields
\be
	\ri( \tk )
		=
			\sqrt[3\,]{\frac{3\.\MB}{4\pi\.\rho}}
		=
			r_{\rm g} \left( \frac{2\.\tk}{r_{\rm g}} \right)^{\!2 / 3},
			\label{eq:ri(t)}
\ee
with the Schwarzschild radius $r_{\rm g} \equiv 2\.G\MB$. For $r > \ri( \tk )$, the kinematic properties of WIMPs are the same as what is obtained for the standard radiation-cosmological history of the Universe, while for $r < \ri( \tk )$, the gravitational attraction of the PBH governs the WIMPs orbiting.

\begin{figure}
\begin{center}
	\includegraphics[width = \linewidth]{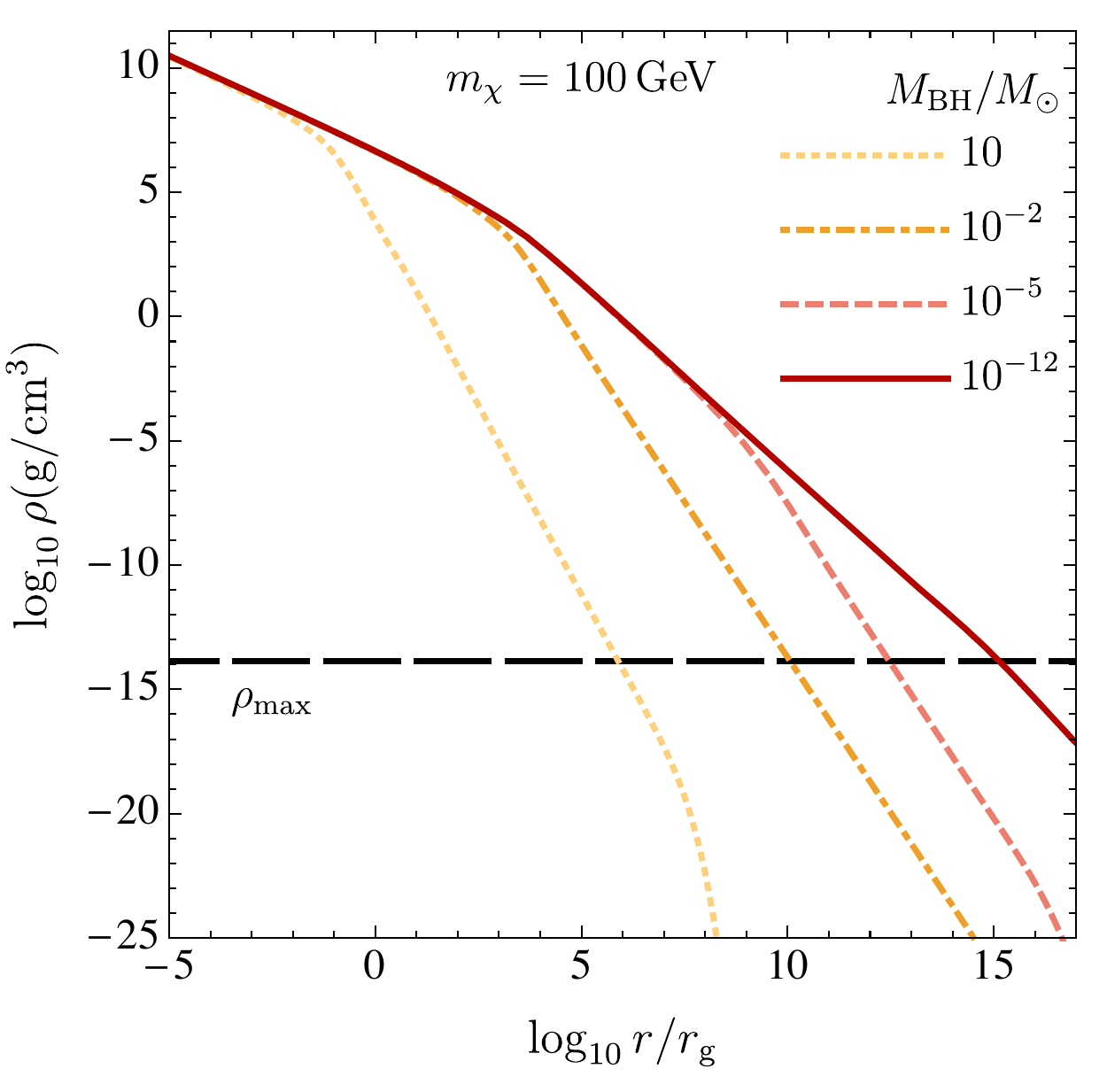}
	\caption{WIMP density profile $\rho( r / r_{\rm g} )$ 
			around a PBH of mass $\MB$
			[\cf~Eq.~\eqref{eq:rho( r )}]. 
			The values of $\MB$ for each curve are
			$\MB = 10\.M_{\odot}$ (yellow dotted curve), 
			$\MB = 10^{-2}\.M_{\odot}$ (orange dot-dashed curve), 
			$\MB = 10^{-5}\.M_{\odot}$ (red dashed curve), 
			and
			$\MB = 10^{-12}\.M_{\odot}$ (maroon solid curve),
			where $M_{\odot}$ is the solar mass. 
			In all cases, the WIMP mass and the velocity-averaged cross-section have been set
			to $m_{\chi} = 100\,{\rm GeV}$ and $\svz = 3 \times 10^{-26}\,{\rm cm}^{3} / {\rm s}$,
			respectively. The horizontal black long-dashed line shows the value of
			$\rho_{\rm max}$ in Eq.~\eqref{eq:MaxRho}.}
	\label{fig:PlotProfile}
	\end{center}
\end{figure}

Now, we focus on a WIMP at position $r_{i}$ and with velocity $\bv$ when the PBH forms at time $t_{i}$. If $\tau_{\rm orb}$ is the period of the WIMP's orbital motion around the PBH, it would spend only a fraction $ 2\,\d t/\tau_{\rm orb}$ at distances between $r$ and $r+\d r$,\footnote{The factor of 2 comes from the fact that the WIMP passes twice by the same radius given the symmetry of the orbit~\cite{Eroshenko:2016yve}.} where $\d t$ is the time it takes for the WIMP to move from  $r$ to $r+\d r$. Then, at any later time $t>t_{i}$, we have the mass relation
\be
	\rho_{\rm bound}( r )\, 4 \pi r^2\d r
		=
			\int\! 4 \pi\d r^{}_{i}\;r_{i}^{2}\,\rho^{}_{i}( r^{}_{i} )
			\int\!\d^{3}\bv\;f_{\Brm}( \bv )\, \frac{2\d t}{\tau_{\rm orb}}
			\; ,
			\label{eq:DensityR}
\ee
which implies that the density of WIMPs in bound orbits around the PBH is given by
\be
	\rho_{\rm bound}( r )
		=
			\frac{1}{r^{2}}
			\int\!\d r^{}_{i}\;r_{i}^{2}\,\rho^{}_{i}( r^{}_{i} )
			\int\!\d^{3}\bv\;f_{\Brm}( \bv )\,\frac{2}{\tau_{\rm orb}}\frac{\d  t}{\d r}
			\; ,
			\label{eq:DensityR}
\ee
where the WIMP velocity distribution $f_{\Brm}( \bv )$, the radial velocity $\d r / \d t$, and the orbital period $\tau_{\rm orb}$ are given in App.~\ref{sec:Derivation-of-the-WIMP-density} in Eqs.~\eqref{eq:BoltzmannDistribution},~\eqref{eq:radialspeed}, and~\eqref{eq:period}, respectively. Reference~\cite{Eroshenko:2016yve} lacks details on the subtleties in the computation of Eq.~\eqref{eq:DensityR}, which deals with the conditions that the WIMP orbit is bound to the PBH. We derive these details in App.~\ref{sec:Derivation-of-the-WIMP-density}.\footnote{It should be noted that our derivation does not take into account general relativistic effects and extreme eccentricities of the WIMP orbits such as those that intersect the PBH. We also work under the assumption that the WIMPs are non-relativistic. Nevertheless, all of these approximations are justified by the fact that the temperature of kinetic decoupling, at which the physics of interest takes place, is much smaller than the WIMP masses considered.}
Accounting for the fact that the present DM density has decreased at least to the value~\cite{Bertone:2005xz}
\be
	\rho_{\rm max}
		=
			\frac{m_{\chi}}{\svz\.t_{0}}
			\; ,
			\label{eq:MaxRho}
\ee
where $t_{0}$ is the age of the Universe, we estimate the present WIMP density profile around a PBH to be
\be
	\rho( r )
		=
			\min\mspace{-2mu}
			\left[
				\rho_{\rm bound}( r ),
				\rho_{\rm max}
			\right]
			\; .
			\label{eq:rho( r )}
\ee
Figure~\ref{fig:PlotProfile} shows the WIMP density profile bound to a PBH as a function of the rescaled radius $x \equiv r / r_{\rm g}$, obtained from Eq.~\eqref{eq:rho( r )} with different values of $\MB$, namely $\MB = 10\.M_{\odot}$, $\MB = 10^{-2}\.M_{\odot}$, $\MB = 10^{-5}\.M_{\odot}$, and $\MB = 10^{-12}\.M_{\odot}$, where $M_{\odot}$ is the solar mass. The horizontal line represents the value of $\rho_{\rm max}$ in Eq.~\eqref{eq:MaxRho} for $m_{\chi} = 100\,{\rm GeV}$ and $\svz = 3 \times 10^{-26}\,{\rm cm}^{3} / {\rm s}$, and thus, we find $\rho_{\rm max} \simeq 1.4 \times 10^{-14}\,{\rm g / cm}^{3}$.

The WIMP density profile for the smallest value of $\MB$ constitutes an envelope to the profiles for the other values and is defined as
\be
	\rho_{0}( r )
		\equiv
			\lim_{\MB\,\to\,0} \rho( r )
			\; .
\ee
We have numerically confirmed that profiles with even smaller values of $\MB$ than $\MB = 10^{-12}\,M_\odot$ converge towards $\rho_{0}( r )$. Increasing the value of $\MB$ from $10^{-12}\.M_{\odot}$ to $10\.M_{\odot}$ and even further, the different profiles decouple from $\rho_{0}( r )$ and cross $\rho_{\rm max}$ at decreasing values of $r$, as is indicated in Fig.~\ref{fig:PlotProfile}. Furthermore, we denote by $\bar{x}$ the ``critical'' value of $x$ at which $\rho_{0}( \bar{x} ) = \rho_{\rm max}$. This corresponds to the ``critical'' PBH mass $\Mb$. Since the WIMP density profile for values below $\rho_{\rm max}$ rapidly goes to zero as a function of $x$, we model the DM density profile around the PBH as
\be
	\rho
		\simeq
			\rho_{\rm max}\,\Theta( \bar{x} - x )
			\; .
			\label{eq:RhoApprox}
\ee
For PBHs with masses $\MB \lesssim \Mb$, the decoupling of the corresponding profiles from $\rho_{0}( r )$ occurs at values of the density below $\rho_{\rm max}$, so that these PBHs share the same constant value of $\bar{x}$, whereas for $\MB \gtrsim \Mb$, $\bar{x}$ is proportional to a power of $\MB$.

\section{Annihilation Signal from Dark-Matter Density Spikes}
\label{sec:Annihilation-Signal-from-Dark--Matter-Density-Spikes}

\noindent In Sec.~\ref{sec:Dark--Matter-Density-Spikes-from-Primordial-Black-Holes}, we have derived the expression for the WIMP density profile around PBHs. Now, we proceed to calculate the expected signal from these overdense regions. The number of WIMP annihilations in the vicinity of a PBH per unit time is given by
\be
	\GB
		=
			\svz\,\int \d^{3} r\;n^{2}( r )
		=
			\frac{4\pi\.\svz}{m_{\chi}^{2}}\,r_{\rm g}^{3}\,\int\!\d x\;x^{2}\rho^{2}( x )
			\; ,
			\label{eq:Gamma}
\ee
where again $x \equiv r / r_{\rm g}$. Based on what we displayed in Fig.~\ref{fig:PlotProfile} and on the parametrization of Eq.~\eqref{eq:RhoApprox}, the decay rate in Eq.~\eqref{eq:Gamma} is
\begin{align}
	\GB
		&\simeq
			\frac{4\pi\.\svz}{m_{\chi}^{2}}\,r_{\rm g}^{3}\,
			\int_{0}^{\bar{x}} \d x\;x^{2}\rho_{\rm max}^{2}
			\notag
			\\
		&=
			\frac{4\pi\.\svz}{3 m_{\chi}^{2}}\,\(2G\MB\)^{3}\,
			\bar{x}^{3} \rho_{\rm max}^{2}
			\; .
			\label{eq:GammaApprox}
\end{align}
Since $\bar{x}$ is constant for $\MB \lesssim \Mb$, we obtain $\GB \propto \MB^{3}$, while the dependence of $\bar{x}$ on $\MB$ for $\MB \gtrsim \Mb$ yields milder relations.

\begin{figure}
\begin{center}
	\includegraphics[width = \linewidth]{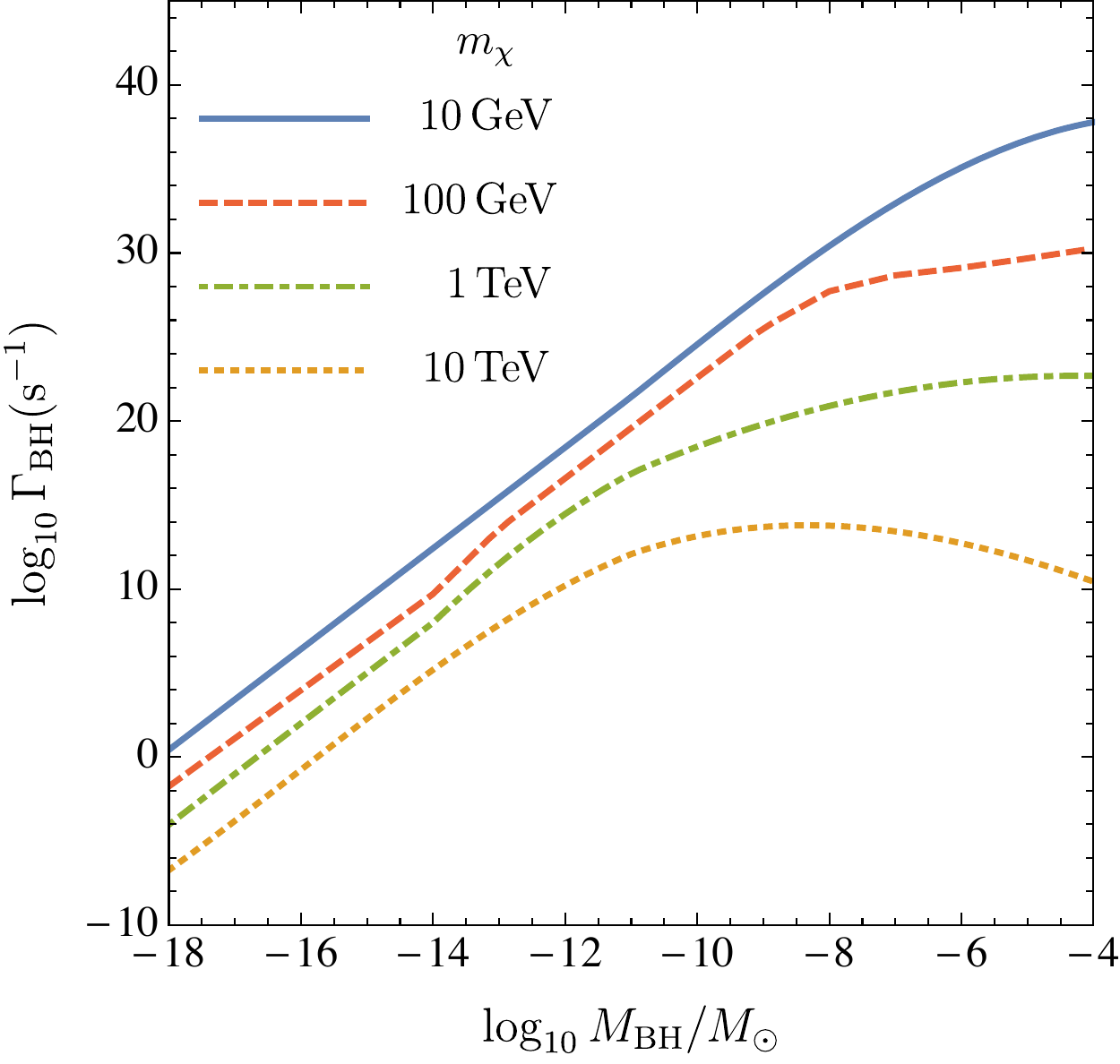}
	\caption{Decay rate $\GB$ as a function of the PBH mass $\MB$
			[see Eq.~\eqref{eq:GammaApprox}] in units of the solar mass $M_{\odot}$
			for different values of the WIMP mass $m_{\chi}$;
			specifically, 
			$m_{\chi} = 10\,{\rm GeV}$ (blue solid curve), 
			$m_{\chi} = 100\,{\rm GeV}$ (red dashed curve), 
			$m_{\chi} = 1\,$TeV (green dot-dashed curve), and
			$m_{\chi} = 10\,$TeV (yellow dotted curve).
			All cases utilize $\svz = 3 \times 10^{-26}\,{\rm cm}^{3} / {\rm s}$,
			and again, assume a sufficient amount of WIMPs.}
	\label{fig:PlotGamma}
	\end{center}
\end{figure}

In Fig.~\ref{fig:PlotGamma}, we show $\GB$ as a function of $\MB$ for different values of $m_{\chi}$, which are chosen as $m_{\chi} = 10\,{\rm GeV}$, $m_{\chi} = 100\,{\rm GeV}$, $m_{\chi} = 1\,{\rm TeV}$, and $m_{\chi} = 10\,{\rm TeV}$. Nevertheless, for small values of $\MB$ (according to Fig.~\ref{fig:PlotGamma}), we have $\GB \propto \MB^{3}$. In general, for the different regimes of $\MB$, the complete power law behavior of $\GB$ is manifest in Fig.~\ref{fig:PlotGamma}. The subscript `BH' reminds us that $\GB$ depends on $\MB$.

The byproducts of WIMPs annihilating around the PBHs contribute to the isotropic flux of gamma rays coming from our Galaxy (``gal'') and the extragalactic (``ex'') components. In fact, the spectrum of gamma rays contains various components like individual sources, the diffuse Galactic emission, and the residual isotropic diffuse gamma-ray background, which comprises the unresolved extragalactic emissions as well as the residual Galactic foregrounds. Therefore, the total expected differential flux of gamma rays is given by~\cite{Cirelli:2012ut}
\be
	\frac{\d \Phi_{\gamma}}{\d E}
		=
			4\pi\.\frac{\d^{2}\Phi_{\gamma}}{\d E\,\d\Omega}\bigg|_{\rm gal}
			+
			\frac{\d\Phi_{\gamma}}{\d E}\bigg|_{\rm ex}
			\; .
			\label{eq:dPhi/dE/dOmega}
\ee
The Galactic component is not isotropic, and hence, the dependence on the differential solid angle $\d\Omega$. However, its minimum value with respect to the direction contributes an irreducible background to the isotropic flux. In the following, we assume that the energy density in the PBHs tracks the DM density profile in both the Galactic and the extragalactic environments as in Eq.~\eqref{eq:Fraction}, i.e.,~$\rho_{\rm BH} = f\.\rDM$.

For a PBH of mass $\MB$ and for a specific annihilation channel, the expected Galactic component of the gamma-ray flux per solid angle is~\cite{Chen:2009uq, Cirelli:2009dv, Cirelli:2012ut}
\be
	\frac{\d^{2}\Phi_{\gamma}}{\d E\,\d\Omega} \bigg|_{\rm gal}
		=
			\frac{f\,(1 - f)^{2}\,\GB}{\MB}\,\frac{1}{4\pi}
			\int_{\rm l.o.s.}\!\d s\;\frac{\d N_{\gamma}}{\d E}\,\rho_{\rm H}(r)
			\; ,
			\label{eq:DiffFluxGalactic}
\ee
where the integral is taken along the line of sight (l.o.s.) $s$, $\d N_{\gamma} / \d E$ is the number of photons $N_\gamma$ produced by the WIMP annihilation channel considered per unit energy $E$, $\rho_{\rm H}(r)$ is the DM distribution in the halo, and we account for the fact that WIMPs contribute a fraction $1 - f$ of the total dark-matter energy density. The coordinate $r$ appearing in $\rho_{\rm H}(r)$ is defined as
\be
	r
		=
			r(s, \psi)
		\equiv
			\sqrt{ r_{\odot}^{2} + s^{2} - 2\.r_{\odot}\.s\.\cos\psi \,}
		\; ,
\ee
where $r_{\odot} = 8.33\,{\rm kpc}$ is the distance from the solar system to the Galactic center, $s$ parametrizes the distance from the solar system along the l.o.s., and $\psi$ is the angle between the direction of observation in the sky and the Galactic center. Similarly, the extragalactic component is~\cite{Chen:2009uq, Cirelli:2009dv, Cirelli:2012ut}
\be
	\frac{\d\Phi_{\gamma}}{\d E}\bigg|_{\rm ex}
		\!\!=
			\frac{f\,(1 - f)^{2}\,\GB}{\MB}\rDM\!
			\int_{0}^{\infty}\!\!\d z\.\frac{\d N_{\gamma}}{\d E}\mspace{-2mu}
			\frac{e^{- \tau_{\rm opt}(z)}}{H(z)}\; ,
			\label{eq:DiffFluxExtra}
\ee
where the optical depth $\tau_{\rm opt}(z)$ accounts for the attenuation of high-energy gamma rays due to their scattering with UV extragalactic photons as a function of the redshift $z$, and $H( z )$ is the Hubble rate as a function $z$.

\section{Results on Constraints on the Primordial Black-Hole Dark-Matter Fraction}
\label{sec:Results}

\noindent Equations~\eqref{eq:DiffFluxGalactic} and~\eqref{eq:DiffFluxExtra} are analogous to what is obtained when considering (a single component) \emph{decaying} DM of mass $m_{\rm DM}$ and decay rate $\Gamma_{\rm DM}$ if we perform the substitution
\be
	\frac{f\,(1 - f)^{2}\,\GB}{\MB}
		=
			\frac{\Gamma_{\rm DM}}{m_{\rm DM}}
			\; .
			\label{eq:analogy}
\ee
Therefore, we can translate the experimental bounds on decaying DM to our parameter spaces.

First, we constrain the PBH physics in the parameter space given by $\MB$ and $f$. We use the results presented in Ref.~\cite{Ando:2015qda}, where the extragalactic gamma-ray background measured by {\sc Fermi} is used to set limits on the DM decay rate $\Gamma_{\rm DM}$ for different decay channels and DM particle masses. Thus, we convert the bounds obtained in Ref.~\cite{Ando:2015qda} to bounds on $f$, given $\GB$ in Eq.~\eqref{eq:GammaApprox} and the analogy in Eq.~\eqref{eq:analogy}. To this end, we focus on the bounds obtained from the $b\bar{b}$ channel, corresponding to the results in Fig.~3.(f) of Ref.~\cite{Ando:2015qda}.

\begin{figure*}
\begin{center}
	\includegraphics[width = \linewidth]{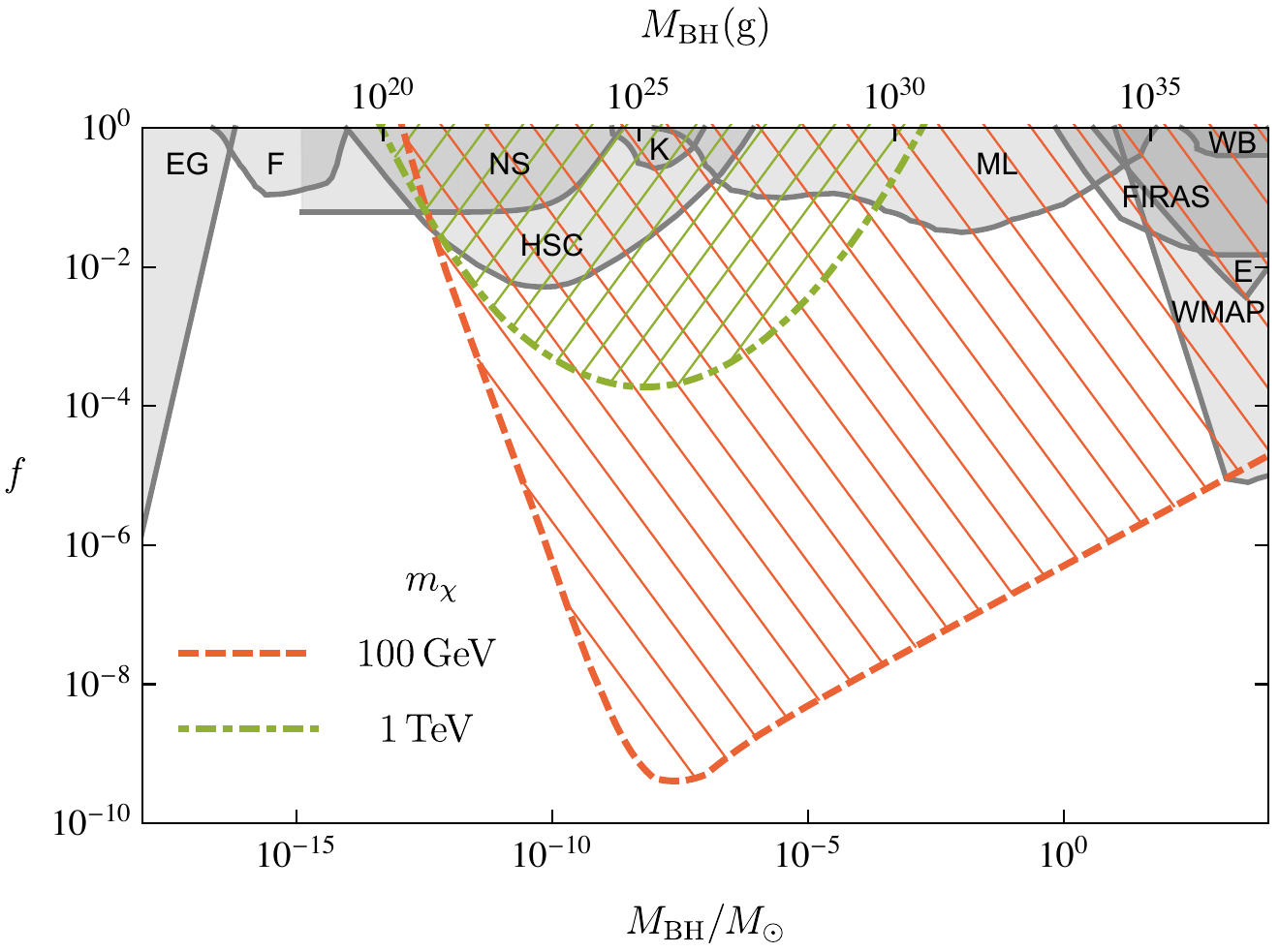}
	\caption{Constraints on the allowed PBH DM fraction $f$
			for a variety of effects associated with PBHs of mass $\MB$
			in units of the solar mass $M_{\odot}$. 
			Here, a monochromatic PBH mass spectrum has been employed.
			The red dashed and green dot-dashed curves are our results, 
			corresponding to 
			$m_{\chi} = 100\,{\rm GeV}$ (red, dotted)
			and
			$m_{\chi} = 1\,{\rm TeV}$ (green, dot-dashed),
			respectively.
			For both cases, $\svz = 3 \times 10^{-26}\,{\rm cm}^{3} / {\rm s}$ has been used.
			Also shown are constraints from 
			extra-Galactic $\gamma$-rays from evaporation (EG) \cite{Carr:2009jm},
			femtolensing of $\gamma$-ray bursts (F) \cite{Barnacka:2012bm}, 
			neutron-star capture (NS) \cite{Capela:2013yf}, 
			microlensing with the {\sc Subaru/HSC} Andromeda observation (HSC)
			\cite{Niikura:2017zjd},
			{\sc Kepler} microlensing of stars (K) \cite{Griest:2013aaa}, 
			{\sc EROS-2} \cite{Tisserand:2006zx} and 
			{\sc OGLE-III} \cite{Wyrzykowski:2011tr} microlensing of stars (ML),
			survival of a star cluster in Eridanus II (E) \cite{Brandt:2016aco}, 
			accretion effects (WMAP and FIRAS) \cite{Ricotti:2007au},
			and disruption of wide binaries (WB) \cite{Monroy-Rodriguez:2014ula}.}
	\label{fig:PlotBounds}
	\end{center}
\end{figure*}

In Fig.~\ref{fig:PlotBounds}, we present the bounds on $f$ as a function of $\MB$ obtained using the analysis on decaying dark matter in Ref.~\cite{Ando:2015qda}. We have used $\svz = 3 \times 10^{-26}\,$cm$^{3}$/s and presented two results, corresponding to $m_{\chi} = 100\,{\rm GeV}$ (red dashed curve) and $m_{\chi} = 1\,{\rm TeV}$ (green dot-dashed curve), respectively. The gray-shaded areas stem from other techniques to constrain the PBH parameter space, see the caption of Fig.~\ref{fig:PlotBounds}. For $m_{\chi} = 100\,{\rm GeV}$, we find that the corresponding bound is, in some regions, several orders of magnitude stronger than \emph{all} currently existing bounds for PBHs with masses $10^{-12}\,M_{\odot} \lesssim \MB \lesssim 10^{4}\,M_{\odot}$.

Second, we convert the bounds on decaying DM to bounds on $f$ in the WIMP parameter space spanned by $m_{\chi}$ and $\svz$ in order to be as model-independent as possible from the point of view of WIMPs. In this case, we fix the PBH parameters to some representative benchmark values, namely $\MB = 10^{-12}\,M_{\odot}$, $\MB = 10^{-5}\,M_{\odot}$, $\MB = 10^{-2}\,M_{\odot}$, and $\MB = 10\,M_{\odot}$. 

\begin{figure*}
\begin{center}
	\includegraphics[width = 0.85\linewidth]{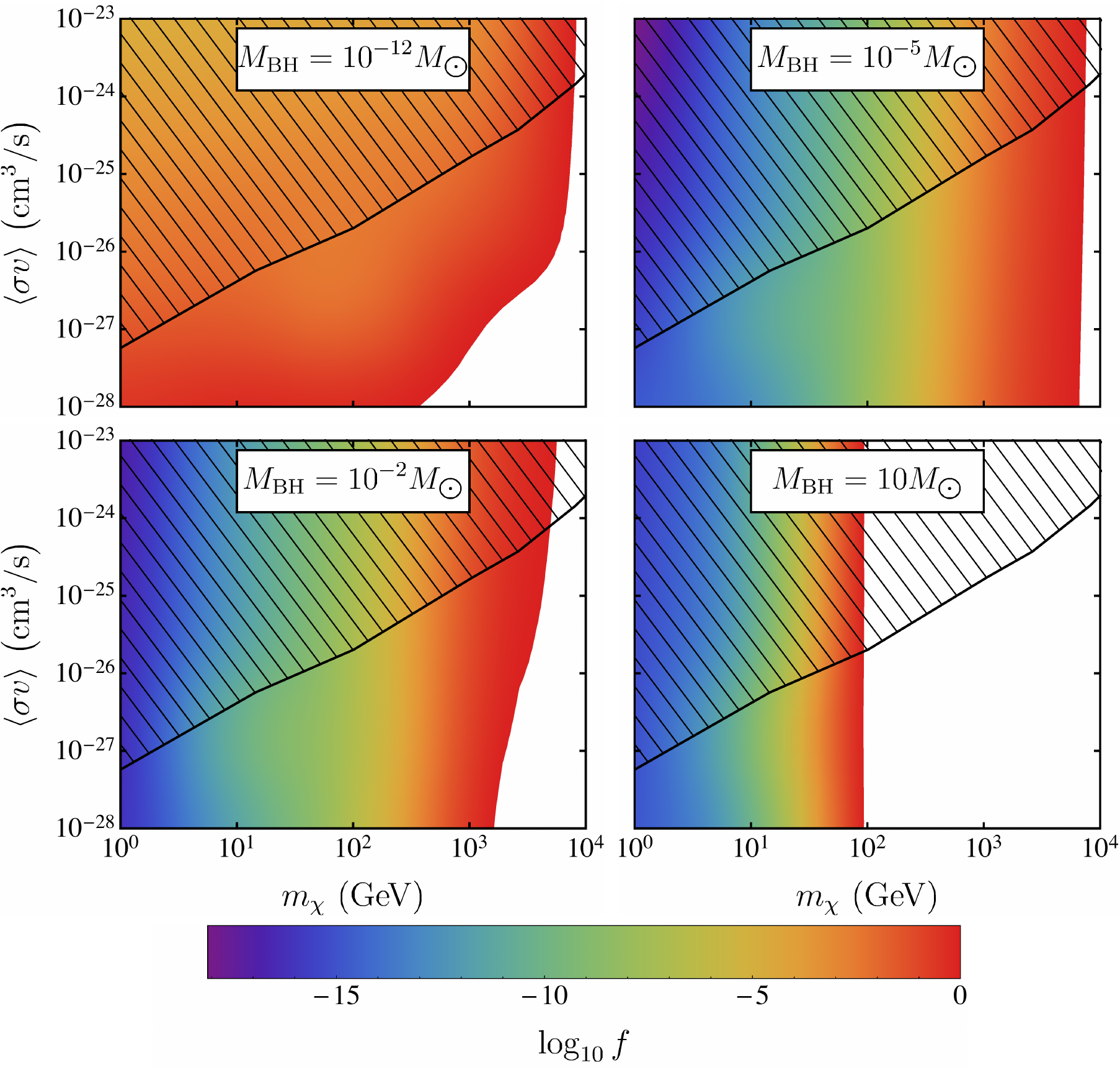}
	\caption{Constraints on the allowed PBH DM fraction $f$ 
			as a function of the WIMP mass $m_{\chi}$
			(in units of ${\rm GeV}$) and the velocity-averaged cross-section $\svz$ 
			(in units of $3 \times 10^{-26}\,{\rm cm}^{3} / \srm$). 
			For each panel, the PBH mass $\MB$ has been fixed:
			$\MB = 10^{-12}\,M_{\odot}$ (upper-left panel),
			$\MB = 10^{-5}\,M_{\odot}$ (upper-right panel),
			$\MB = 10^{-2}\,M_{\odot}$ (lower-left panel), and
			$\MB = 10\,M_{\odot}$ (lower-right panel).
			As before, a monochromatic PBH mass spectrum has been used.
			The white region in each panel represents values of $f > 1$, which are excluded,
			whereas the black solid curve in each panel shows the constraint from {\sc Fermi} 
			and {\sc MAGIC} telescopes for the $b\bar{b}$ channel \cite{Ahnen:2016qkx},
			and the black hatched region (valid for $f \ll 1$) marks 
			the corresponding values of $m_{\chi}$ and $\svz$, which are excluded.}
	\label{fig:PlotBounds-sigmav-mchi}
	\end{center}
\end{figure*}

In Fig.~\ref{fig:PlotBounds-sigmav-mchi}, we display density plots for the PBH fraction $f$ as a function of $m_{\chi}$ and $\svz$ for the four above-mentioned values of $\MB$. The colored regions of these plots represent $f$ with the color scale indicating the value of $\log_{10}f$. White regions show areas in which the value of $f>1$ and are therefore excluded. The hatched regions mark the areas of the WIMP parameter space that are excluded by the search of gamma rays from DM annihilation in dwarf satellite galaxies coming from the combined analysis of the {\sc Fermi} and {\sc MAGIC} telescopes~\cite{Ahnen:2016qkx} for the $b\bar{b}$ channel. This bound assumes that WIMPs account for the total DM of the Universe and is therefore only valid for $f \ll 1$, otherwise it should be properly rescaled for the considered value of $f$. We show it to illustrate the interplay between the WIMP indirect-detection bounds and $f$. Figure \ref{fig:PlotBounds-sigmav-mchi} provides the maximal value of $f$ for each WIMP parameter pair $m_{\chi}$ and $\svz$.

\section{Conclusions}
\label{sec:Conclusions}

\noindent In this work, we have derived constraints on mixed DM scenarios consisting of PBHs and WIMPs. Here, the PBHs efficiently accrete the WIMPs, leading to spikes in their DM density profile. We have precisely calculated this profile (see Fig.~\ref{fig:PlotProfile}) as well as the associated enhanced WIMP annihilation signal (see Fig.~\ref{fig:PlotGamma}). These are shown to potentially provide strong constraints on the allowed PBH DM fraction $f$. In particular, we have demonstrated that experimental knowledge on WIMP DM may be used to significantly constrain both the PBH and WIMP parameter spaces. Our results are summarized in Figs.~\ref{fig:PlotBounds} and \ref{fig:PlotBounds-sigmav-mchi}.

We should stress that the exclusion limits derived in this work are for the mixed scenario of PBH and WIMP DM. In particular, these constraints by no means restrict the abundance of PBHs for DM particles of much larger masses and/or smaller annihilation cross-sections.

\begin{acknowledgments}

\noindent We are grateful to M.~Taoso for useful discussions on {\sc Fermi} bounds. 
S.B.~thanks the ``Roland Gustafssons Stiftelse f{\"o}r teoretisk fysik'' for financial support. F.K.~and L.V.~acknowledge support by the Swedish Research Council (Vetenskapsr\r{a}det) through contract No.~638-2013-8993 and the Oskar Klein Centre for Cosmoparticle Physics. T.O.~acknowledges support by the Swedish Research Council (Vetenskapsr\r{a}det) through contract No.~2017-03934 and the KTH Royal Institute of Technology for a sabbatical period at the University of Iceland.

\end{acknowledgments}

\appendix

\section{Primordial Black-Hole Formation}
\label{sec:Primordial-Black--Hole-Formation}

\noindent When re-entering the cosmological horizon, a density perturbation collapses to a PBH if it is larger than a certain medium- and shape-dependent threshold. Here, the best studied (and highly idealized) case is that of a collapse of spherical Gaussian overdensities within the epoch of radiation domination. The vast majority of the literature assumes rapid dynamics, leading to a black hole with a mass proportional to the horizon mass, and hence, a monochromatic mass spectrum. However, it can be shown that even if the initial density spectrum was monochromatic, the phenomenon of critical collapse \cite{Niemeyer:1997mt} will inevitably lead to a PBH mass spectrum which is spread out, shifted towards lower masses and lowered, leading to potentially large effects (\cf~Ref.~\cite{Kuhnel:2015vtw}). Under the assumption of spherical symmetry, it has been shown \cite{Choptuik:1992jv, Koike:1995jm, Niemeyer:1999ak, Gundlach:1999cu, Gundlach:2002sx} that the functional dependence of the PBH mass $\MB$ on the density contrast $\delta$ follows the critical scaling relation
\be
	\MB
		=
			k\.M_{\rm H}\.
			\big(
				\delta
				-
				\delta_{\crm}
			\big)^{\gamma_{\crm}}
			\qq
			\hbox{for $\delta > \delta_{\crm}$}
			\; ,
			\label{eq:M-delta-scaling}
\ee 
where $M_{\rm H}$ is the mass contained in a Hubble patch at a given time $t$, namely
\be
	M_{\rm H}
		\simeq
			\frac{ c^{3}\, t }{ G }
		\sim
			10^{15}
			\left(
				\frac{t}{10^{-23}\,\srm}
			\right)
			\grm
			\; .
			\label{eq:Moft}
\ee
In Eq.~\eqref{eq:M-delta-scaling}, the constant $k$, the threshold $\delta_{\crm}$, and the critical exponent $\gamma_{\crm}$ all depend on the nature of the fluid containing the overdensity $\delta$ at horizon-crossing \cite{Musco:2012au}. In radiation-dominated models, which are the focus of this paper, repeated studies have shown that $\gamma_{\crm} \simeq 0.36$ \cite{Koike:1995jm, Niemeyer:1999ak, Musco:2004ak, Musco:2008hv, Musco:2012au} and $\delta_{\rm c} \simeq 0.45$ \cite{Musco:2004ak, Musco:2008hv, Musco:2012au}. In accordance with Ref.~\cite{Niemeyer:1997mt}, we set $k = 3.3$. Precise numerical computations \cite{Musco:2004ak, Musco:2008hv, Musco:2012au} have confirmed the above scaling law, which has been shown to apply over more than ten orders of magnitude in density contrast \cite{Musco:2008hv}. Applying the Press--Schechter formalism \cite{1974ApJ...187..425P} for spherical collapse and assuming a Gaussian perturbation profile, one can express the ratio $\beta$ of the PBH energy density to the total energy density at the time of PBH \emph{formation} as
\be
\beta
		\approx
			k\.\sigma^{2\gamma_{\crm}}\,
			{\rm erfc}
			\bigg(
				\frac{ \delta_{\crm} }{ \sqrt{2\,}\.\sigma }
			\bigg)
			\; ,
			\label{eq:BetaNormalisation}
\ee
which holds for $\sigma \ll \delta_{\crm}$ with $\sigma$ being the variance of the primordial power spectrum of density perturbations generated by the model of inflation. From the above specified $\beta$, we can express the PBH DM fraction $f$ via
\be
	f
		=
			\frac{\beta^{\rm eq}}{\Omega_{\rm DM}^{\rm eq}}
		\approx
			2.4\.\beta^{\rm eq}
			\; ,
			\label{eq:Fraction-2}
\ee
where $\Omega_{\rm DM}^{\rm eq} \approx 0.42$ and $\beta^{\rm eq}$ are the DM density and the PBH mass fraction at matter-radiation equality, respectively.\footnote{When using critical collapse instead of the horizon-mass approximation, at each instance of time $t$ of horizon re-entry of a given mode, we have a to deal with an extended PBH mass spectrum. Hence, one looses the one-to-one correspondence between $M_{\rm BH}$ and $t$. In order to evaluate $\beta$ (or $f$) at a given time, one needs to take into account the scaling of the PBH population, behaving as matter in radiation domination, leading to different amplifications for different modes (see Ref.~\cite{Kuhnel:2015vtw} for details).}

In the main part of this work, we derive the flux $\Phi_{\gamma} |_{\MB}$ of gamma rays coming from WIMP annihilation in halos around the PBHs assuming a certain, fixed $\MB$. However, for a given extended mass distribution $f( \MB )$, specifying the number of PBHs per unit mass, the final constraint can be easily derived from these results. This utilizes the following weighting: Let $I_{\rm bin} \in \Nbb$ denote the number of mass bins $\{ [ M_{{\rm BH}, i},\.M_{{\rm BH}, i + 1} ],\, 1 \leq i \leq I_{\rm bin}\}$. Then, we have
\be
	\Phi_{\gamma} \big|_{\rm total}
		\approx
			\sum_{i = 1}^{I_{\rm bin}}\.\Phi_{\gamma} \big|_{{\MB}}^{(i)}
			\int_{M_{{\rm BH}, i}}^{M_{{\rm BH}, i + 1}}\!\d \MB\;
			\frac{ \d \ln f( \MB ) }{ \d \MB}
			\; ,
			\label{eq:ConstraintNSCapture}
\ee
where $\Phi_{\gamma} \big|_{\MB}^{(i)}$ is evaluated for a mass within each bin $i$ and the bin number $I_{\rm bin}$ should be chosen such that, to the precision sought, it does not matter where exactly in each bin $i$ the quantity $\Phi_{\gamma} \big|_{\MB}^{(i)}$ is evaluated (\cf~Ref.~\cite{Kuhnel:2017pwq}).

\section{Kinetic Decoupling}
\label{sec:Kinetic-Decoupling}

\noindent We consider the single-particle Boltzmann equation, neglecting other effects related to the medium and restricting to Boltzmann statistics. After the chemical decoup\-ling has occurred at temperature $T_{\rm chem} \sim m_{\chi} / 20$, the relevant Boltzmann collision equation for the phase space density $f_{\Brm}$ reads \cite{Bernstein:1985, Bernstein:1988}
\be
	E
	\left(
		\frac{\partial}{\partial t}
		-
		H\,\bp\cdot\!\nabla_{\! p}
	\right)
	f_{\Brm}( \bp ) 
		=
			C_{\rm el}[ f_{\Brm} ]
			\; .
			\label{eq:BoltzmannF}
\ee
Here, the elastic collision rate is given by~\cite{Bringmann:2006mu, Bringmann:2009vf}
\be
	C_{\rm el}[ f_{\Brm} ]
		=
			2\.E\,\gamma( T )
			\left(
				m_{\chi}\.T\.\nabla_{\!p}^{2}
				+
				\bp\cdot\!\nabla_{\!p}
				+
				3
			\right)
			f_{\Brm}( \bp )
			\; ,
			\label{eq:Cel[f]}
\ee
where $\gamma(T)$ is the momentum relaxation rate. Introducing the number density and the kinetic temperature of $\chi$ as
\be
	n_{\chi}
		=
			\int\!\frac{\d^{3}\bp}{( 2\pi )^{3}}\;f_{\Brm}( \bp )
			\; ,
	\mspace{10mu}
	T_{\chi}
		=
			\frac{1}{3\.m_{\chi}\.n_{\chi}}
			\int\!\frac{\d^{3}\bp}{( 2\pi )^{3}}\;
			p^{2}\.f_{\Brm}( \bp )
			\; , 
			\label{eq:NChiTChi}
\ee
respectively, the expression for $T_{\chi}$ is obtained by integrating Eq.~\eqref{eq:BoltzmannF} by $\int\mspace{-2mu}\d^{3}\bp\;p^{2} / [ (2\pi)^{3} E ]$, obtaining
\be
	\dot{T}_{\chi} + 2\.H\.T_{\chi}
	-
	\gamma( T )
	\left(
		T - T_{\chi}
	\right)
		=
			0
			\; .
			\label{eq:BoltzmannT}
\ee
The temperature of kinetic decoupling is defined as~\cite{Bringmann:2006mu, Bringmann:2009vf}
\be
	\Tk
		=
			\frac{T^{2}}{T_{\chi}}\bigg|_{T\.\to\.0}
		=
			\frac{m_{\chi}}{\Gamma( 3 / 4 )} \left( \frac{\alpha\.m_{\chi}}{\MP} \right)^{\!\! 1 / 4}
			\; ,
			\label{eq:Tk}
\ee
where $\Gamma(\cdot)$ is the gamma function. This expression of $\Tk$ coincides, up to a numerical factor, with other definitions available in the literature~\cite{Visinelli:2015eka}.

\section{Derivation of the WIMP Density}
\label{sec:Derivation-of-the-WIMP-density}

\noindent In deriving the WIMP density, our starting point is~\cite{Eroshenko:2016yve}
\be
	\rho( r )
		=
			\frac{1}{r^{2}}\!\int \d r_{i}\;
			r_{i}^{2}\,\rho_{i}( r_{i} )
			\int\!\d^{3}\bv\;
			f_{\Brm}( \bv )\,\frac{2\,\d t / \d r}{\tau_{\rm orb}}
			\; ,
			\label{eq:DensityRa}
\ee
which is exactly the same equation as Eq.~\eqref{eq:DensityR}. The specific expressions for $\tau_{\rm orb}$, $\d t / \d r$, and $f_{\Brm}( \bv )$ will be specified below in Eqs.~\eqref{eq:period},~\eqref{eq:radialspeed}, and~\eqref{eq:BoltzmannDistribution}, respectively. The energy of a particle of mass $m_{\chi}$ at position $r_{i}$ and with velocity $v_{i}$ is given by
\be
	E
		=
			\frac{m_{\chi}}{2}\.\bv^{2}
			-
			\frac{G\.\MB\,m_{\chi}}{r_{i}} 
		=
			-
			\frac{m_{\chi}}{2}
			\left(
				\frac{1}{r_{i}} - \beta_{i}^{2}
			\right)
			,
			\label{eq:E}
\ee
where in the last equality we have expressed the velocity in terms of $\beta_{i} \equiv v_{i} / c$ and the radius $r_{i}$ in terms of $r_{\rm g}$. We have then set $c = r_{\rm g} = 1$. Requiring that the orbit is bound gives $E \leq 0$ or $r_{i} < 1 / \beta_{i}^{2}$. The orbital period and the inverse radial velocity are
\be
	\tau_{\rm orb}
		=
			2\pi\.G \MB
			\left(
				\frac{m_{\chi}}{2\.| E |}
			\right)^{\!3 / 2}\!
		= 
			\pi\mspace{-1mu}
			\left(
				\frac{1}{r_{i}}
				-
			\beta_{i}^{2}
			\right)^{\!-3 / 2} 
			\label{eq:period}
\ee
and
\begin{subequations}
\begin{align}
	\frac{\d t}{\d r}
		&=
			\left[
				\frac{2}{m_{\chi}}
				\big[
					E - U( r )
				\big]
				-
				\left(
					\frac{l}{m_{\chi} r}
				\right)^{\mspace{-4mu} 2}\.
			\right]^{-1 / 2}
			\notag
			\\
		&=
			\left[
			\beta_{i}^{2} - \frac{1}{r_{i}} + \frac{1}{r}
				-
				\left(
					\frac{r_{i}\.\beta_{i}}{r}
				\right)^{\mspace{-4mu} 2}\!
				\left(
					1 - y^{2}
				\right)
			\right]^{-1 / 2}
			\; ,
			\label{eq:radialspeed}
\end{align}
respectively. 
Above, $y = \cos\theta$, the angular momentum $l$ is given by $l = m_{\chi}\.r_{i}\.v_{i}\.\sin\theta$, and we have expressed all radial quantities in units of $r_{\rm g}$. We rewrite this expression as
\begin{align}
	\frac{\d t}{\d r}
		&=
			\frac{r}{r_{i}\.\beta_{i}}
			\left[
				\left(
					\frac{r}{r_{i}\.\beta_{i}}
				\right)^{\!2}\!
				\left(
				\beta_{i}^{2} - \frac{1}{r_{i}} + \frac{1}{r}
				\right) 
				-
				\left(
					1 - y^{2}
				\right)
			\right]^{-1 / 2}
			\notag
			\\
		&=
			\frac{r}{r_{i}\.\beta_{i}}
			\frac{1}{\sqrt{y^{2} - y_{m}^{2}\,}}
			\; ,
			\label{eq:radialspeed1}
\end{align}
\end{subequations}
where
\be
	y_{m}
		\equiv
			\sqrt{1
				+
				\left(
					\frac{r}{r_{i}\.\beta_{i}}
				\right)^{\!2}\!
				\left(
					\frac{1}{r_{i}} - \beta_{i}^{2} - \frac{1}{r}
				\right)
			\,}
			\; .
			\label{eq:ym}
\ee
The condition that the orbit is confined between the aphelion and the perihelion implies that $y^{2} > y_{m}^{2}$~\cite{Eroshenko:2016yve}. Inserting Eqs.~\eqref{eq:period} and~\eqref{eq:radialspeed1} into Eq.~\eqref{eq:DensityRa} and writing $\d^{3}\bv = 2\pi\.\beta_{i}^{2}\.\d \beta_{i}\,\d y$, the expression for the WIMP density reduces to
\begin{align}
	\rho( r )
		&=
			\frac{8}{r}\int_{0}^{\infty}\!\d\beta_{i}\;\beta_{i}\.f_{\Brm}( \beta_{i} )
			\,\int_{0}^{\infty}\!\d r_{i}\;r_{i}\,\rho_{i}( r_{i} )\;
			\notag
			\\
	&\phantom{=\;} \times
			\left(
				\frac{1}{r_{i}}
				-
			\beta_{i}^{2}
			\right)^{\!3 / 2}
			\int_{y_{m}}^{1}\!\frac{\d y}{\sqrt{y^{2} - y_{m}^{2}\,}}
			\; .
			\label{eq:DensityRSimplified}
\end{align}
Here, we explore the parameter space further in order to perform the integration. Since $0 \leq y_{m}^{2} \leq 1$, we obtain
\be
	0
		\leq
			1
			+
			\left(
				\frac{r}{r_{i}\.\beta_{i}}
			\right)^{\!2}\!
			\left(
				\frac{1}{r_{i}} - \beta_{i}^{2} - \frac{1}{r}
			\right)
		\leq
			1\; .
			\label{eq:Condition}
\ee
The upper bound leads to
\be
	r_{i}
		\geq
			\frac{r}{1 + r\.\beta_{i}^{2}}
			\; ,
			\label{eq:UpperCondition}
\ee
whereas the lower bound gives
\be
	1
	+
	\left(
		\frac{r}{r_{i}\.\beta_{i}}
	\right)^{\!2}\!
	\left(
		\frac{1}{r_{i}} - \beta_{i}^{2} - \frac{1}{r}
	\right)
		\geq
			0
			\; ,
			\label{eq:LowerCondition}
\ee
and the equality has the two positive roots $r_{i, 1}= r$ and $r_{i, 2} = r\,\big\{ [ 1 + 4 / ( r\.\beta_{i}^{2} ) ]^{1 / 2} - 1 \big\} / 2$. Defining
\be
	r_{+}
		\equiv 
				\max( r_{i, 1}, r_{i, 2} )
				\; ,
				\q
	r_{-}
		\equiv
				\min( r_{i, 1}, r_{i, 2} )
				\; ,
				\label{eq:RPlusAndRMinus}
\ee
the condition in Eq.~\eqref{eq:LowerCondition} is satisfied for $r_{i} \leq r_{-}$ or $r_{i} \geq r_{+}$. Finally, setting $x_{m} \equiv \sqrt{1 - y_{m}^{2}\,}$, Eq.~\eqref{eq:DensityRSimplified} reads
\begin{align}
	\rho( r )
		&=
			\frac{8}{r}\int_{0}^{\infty}\!\d\beta_{i}\;\beta_{i} \,f_{\Brm}( \beta_{i} )
			\int_{0}^{\infty}\!\d r_{i}\;r_{i}\,\rho_{i}( r_{i} )
			\notag
			\\
		&\phantom{=\;}\times
			\left(
				\frac{1}{r_{i}}
				-
			\beta_{i}^{2}
			\right)^{\!3 / 2}
			\ln\frac{1 + x_{m}}{y_{m}}\.\Theta
			\; ,
			\label{eq:DensityR1}
\end{align}
where $\Theta$ is defined in terms of the Heaviside function $\theta( \cdot )$ as
\begin{align}
	\Theta
		&\equiv
			\theta\!
			\left(
				\frac{1}{\beta_{i}^{2}} - r_{i}
			\right)
			\theta\!
			\left(
				r_{i} - \frac{r}{1 + r \beta_{i}^{2}}
			\right)
			\notag
			\\
		&\phantom{=\;\.} \times
			\big[\.
				\theta\mspace{-1mu}
				\left(
					r_{-} - r_{i}
				\right)
				+
				\theta\mspace{-1mu}
				\left(
					r_{i} - r_{+}
				\right)
			\big]
			\; .
			\label{eq:Theta}
\end{align}
Here, we use the Maxwell--Boltzmann distribution
\be
	f_{\Brm}( \beta_{i} )
		=
			\frac{1}{( 2\pi\.\bar{\sigma}^{2} )^{3 / 2}}\.
			\exp\!
			\left(
				-
				\frac{\beta_{i}^{2}}{2\.\bar{\sigma}^{2}}
			\right)
			\; ,
			\label{eq:BoltzmannDistribution}
\ee
where $\bar{\sigma} \equiv \sqrt{T / m_{\chi}\,}$. Furthermore, the initial energy density and the temperature at radius $r_{i}$ are given by (cf.~the discussion in Sec.~3 in Ref.~\cite{Eroshenko:2016yve})
\begin{subequations}
\begin{align}
	\rho_{i}( r_{i} )
		&=
			\rk\,\theta\!
			\left(
				\ri - r_{i}
			\right)
			\notag
			\\
		&\phantom{=\;}
			+
			\rk
			\left(
				\frac{\ri}{r_{i}}
			\right)^{\mspace{-4mu}9 / 4}\,
			\theta\!
			\left(
				r_{i} - \ri
			\right)
			\. ,
			\label{eq:rhoi(ri)}
			\\
	T( r_{i} )
		&=
			\Tk\,\theta\!
			\left(
				\ri - r_{i}
			\right)
			\notag
			\\
		&\phantom{=\;}
			+
			\Tk
			\left(
				\frac{\ri}{r_{i}}
			\right)^{\mspace{-4mu}3 / 2}\,
			\theta\!
			\left(
				r_{i} - \ri
			\right)
			\. ,
			\label{eq:T(ri)}
\end{align}
\end{subequations}
respectively.


\end{document}

%% file: header.tex
\usepackage{jcappub}
\usepackage{amsmath}	
\usepackage{amsthm}	
\usepackage{amssymb}	
\usepackage{datetime}
\usepackage{graphicx}
\usepackage{verbatim}
\usepackage{bm}
\usepackage{soul}

\definecolor{grey}{rgb}{0.4,0.4,0.4}
\definecolor{dullmagenta}{rgb}{0.4,0,0.4}
\definecolor{darkblue}{rgb}{0,0,0.4}
\definecolor{midblue}{rgb}{0,0,0.5}
\definecolor{midred}{rgb}{0.5,0,0}
\definecolor{orange}{rgb}{1,0.5,0}
\definecolor{lightbrown}{rgb}{0.75,0.5,0.25}
\definecolor{tan}{cmyk}{0.14,0.42,0.56,0}
\definecolor{djunglegreen}{cmyk}{0.99,0,0.52,0}
\definecolor{lightgreen}{rgb}{0,1,0}
\definecolor{olivegreen}{cmyk}{0.64,0,0.95,0.40}
\definecolor{midgreen}{rgb}{0.0,0.675,0.0}
\definecolor{darkgreen}{rgb}{0,0.5,0}

\newcommand{\q}{\quad}
\newcommand{\qq}{\qquad}

\renewcommand{\.}{\hspace{0.5mm}}

\newcommand{\Brm}{\ensuremath{\mathrm{B}}}

\newcommand{\crm}{\ensuremath{\mathrm{c}}}

\newcommand{\grm}{\ensuremath{\mathrm{g}}}

\newcommand{\srm}{\ensuremath{\mathrm{s}}}

\newcommand{\Nbb}{\ensuremath{\mathbb{N}}}

\renewcommand{\d}{\ensuremath{\mathrm{d}}}

\newcommand{\cf}{cf.}

\newcommand{\be}{\begin{equation}}
\newcommand{\ee}{\end{equation}}
\newcommand{\bea}{\begin{eqnarray}}
\newcommand{\eea}{\end{eqnarray}}
\renewcommand\({\left(}
\renewcommand\){\right)}

\let\baraccent=\= 
\renewcommand{\=}[1]{\stackrel{#1}{=}} 

\theoremstyle{definition}

\theoremstyle{remark}

\smallskipamount

\settimeformat{ampmtime}

\usepackage{float}

\usepackage{epstopdf}
\usepackage{hyperref}
\hypersetup{
     colorlinks   = true,
     citecolor    = blue
}

\newcommand{\bp}{\bm{p}}
\newcommand{\bv}{\bm{v}_{i}}
\newcommand{\ri}{r_{\rm infl}}
\newcommand{\Mb}{\overline{M}_{\rm BH}}
\newcommand{\MB}{M_{\rm BH}}
\newcommand{\MP}{M_{\rm Pl}}
\newcommand{\svz}{\langle \sigma v \rangle}
\newcommand{\rk}{\rho_{\rm KD}}
\newcommand{\rDM}{\rho_{\rm DM}}
\newcommand{\rW}{\rho_{\rm WIMP}}

\newcommand{\Tk}{T_{\rm KD}}
\newcommand{\tk}{t_{\rm KD}}
\newcommand{\GB}{\Gamma_{\rm BH}}

